\documentclass[9pt,twocolumn,twoside]{pnas-new}
\setboolean{displaywatermark}{false}
\usepackage{multicol}
\usepackage{subfig}
\usepackage{hyperref,comment} 

\templatetype{pnasresearcharticle}

\title{Interfacial Cavitation}

\author[a,1]{T. Henzel}
\author[b,1]{J. Nijjer}
\author[c,1]{S. Chockalingam}
\author[d]{H. Wahdat}
\author[d]{A.J.  Crosby}
\author[b,e,*]{J. Yan}
\author[a,f,*]{T. Cohen}

\affil[a]{Department of Civil and Environmental Engineering, Massachusetts Institute of Technology, Cambridge, MA 02139}
\affil[b]{Molecular, Cellular and Developmental Biology , Yale University, New Haven, CT 06520}
\affil[c]{Department of Aeronautics and Astronautics, Massachusetts Institute of Technology, Cambridge, MA 02139}
\affil[d]{Polymer Science and Engineering Department, University of Massachusetts Amherst, Amherst, MA 01003, USA. }
\affil[e]{Quantitative biology Institute, Yale, New Haven, CT 06520}
\affil[f]{Department of Mechanical Engineering, Massachusetts Institute of Technology, Cambridge, MA 02139}

\leadauthor{Henzel} 

\begin{abstract} \vspace{-3mm}
Cavitation has long been recognized as a crucial predictor, or precursor, to the ultimate failure of various materials, ranging from ductile metals to soft and biological materials. 
Traditionally, cavitation in solids is defined as an unstable expansion of a void or a defect within a material. The critical applied load needed to trigger this instability -- the critical pressure -- is a lengthscale independent material property and has been predicted by numerous theoretical studies for a breadth of constitutive models.  While these studies usually  assume that cavitation initiates from  defects in the bulk of an otherwise homogeneous medium, an alternative and potentially more ubiquitous scenario can occur if the defects are found at interfaces between two distinct media within the body. Such interfaces are becoming increasingly common in modern materials with the use of multi-material composites  and layer-by-layer additive manufacturing methods.  
However, a criterion to determine the threshold for interfacial failure, in analogy to the bulk cavitation limit, has yet to be reported. In this work we fill this gap. Our theoretical  model captures a lengthscale independent  limit for interfacial cavitation, and is shown to agree with our  observations at two distinct lengthscales, via two different experimental systems. To further understand the competition between the two cavitation modes (bulk versus interface)  we  expand our investigation beyond the elastic response to understand the ensuing unstable propagation of delamination at the interface. A phase diagram summarizes these results, showing regimes in which  interfacial failure becomes the dominant mechanism. \vspace{-10mm}
\end{abstract}

\dates{$^*$ To whom correspondence should be addressed: T. Cohen  (talco@mit.edu); J. Yan (jing.yan@yale.edu)\\$^1$ These authors contributed equally. }

\begin{document}

\maketitle
\thispagestyle{firststyle}
\ifthenelse{\boolean{shortarticle}}{\ifthenelse{\boolean{singlecolumn}}{\abscontentformatted}{\abscontent}}{}


\dropcap{T}he genesis of the study of cavitation dates back to the seminal work of  Rayleigh in 1917 \cite{rayleigh1917viii}\footnote{Rayleigh mentions the earlier work by Besant \cite{besant1859treatise}, which considered  the same problem but did not resolve the internal pressure.}. Concerned with the growth and subsequent collapse of bubbles in water, Rayleigh proposed a simple model that estimates the internal pressure in a collapsing  spherical cavity. Since then, the violent collapse of bubbles that form near solid surfaces has been studied extensively. 

Perhaps the earliest study of cavitation in solids dates back to the work of Bishop, Hill and Mott in 1945 \cite{bishop1945theory}, which aimed to obtain theoretical predictions for interpretation of indentation and hardness tests in ductile metals. They argued that the maximum resisting pressure attained in the indentation process can be well estimated by the pressure required to expand a cavity indefinitely. Though it is not obvious that a constant finite pressure can induce indefinite expansion, their theory predicted that such an asymptotic pressure exists and is lengthscale independent; it is thus a material property - the \textit{cavitation pressure}. The notion that this material property is useful to determine the ability of a solid to sustain loads has since been extended beyond indentation. By now, it is well established as a criteria for onset of ductile fracture \cite{puttick1959ductile,ashby1966work,tanaka1970cavity,le1981model,tvergaard1982ductile,kassner2003creep}, 
it has recently been indicated as an underlying mechanism of failure in brittle materials\cite{murali2011atomic,shen2021observation}, and it serves as a measure for estimation of static and dynamic penetration and perforation processes \cite{chen2002deep,masri2005dynamic,forrestal2008penetration,borvik2009perforation,cohen2010ballistic}. 
Additionally, in recent years, the cavitation pressure has been pivotal in modern methods to measure properties of soft and biological materials \cite{crosby2011blowing,zimberlin2007cavitation,zimberlin2010cavitation,raayai2019volume,chockalingam2021probing},
 and its use for predicting the failure of rubbers has been subjected to an ongoing debate \cite{gent1990cavitation,lefevre2015cavitation,poulain2017damage,raayai2019intimate}. Moreover, the long-established neo-Hookean cavitation pressure, $p_{bc}/\mu=5/2$, for incompressible materials with shear modulus $\mu$, has become foundational in explaining chemical and biological processes in which cavities can form spontaneously inside the material\cite{goriely2010elastic,kothari2020effect,vidal2021cavitation,kothari2022crucial}.

\begin{figure*}[h]
\centering
\includegraphics[width=0.9\linewidth]{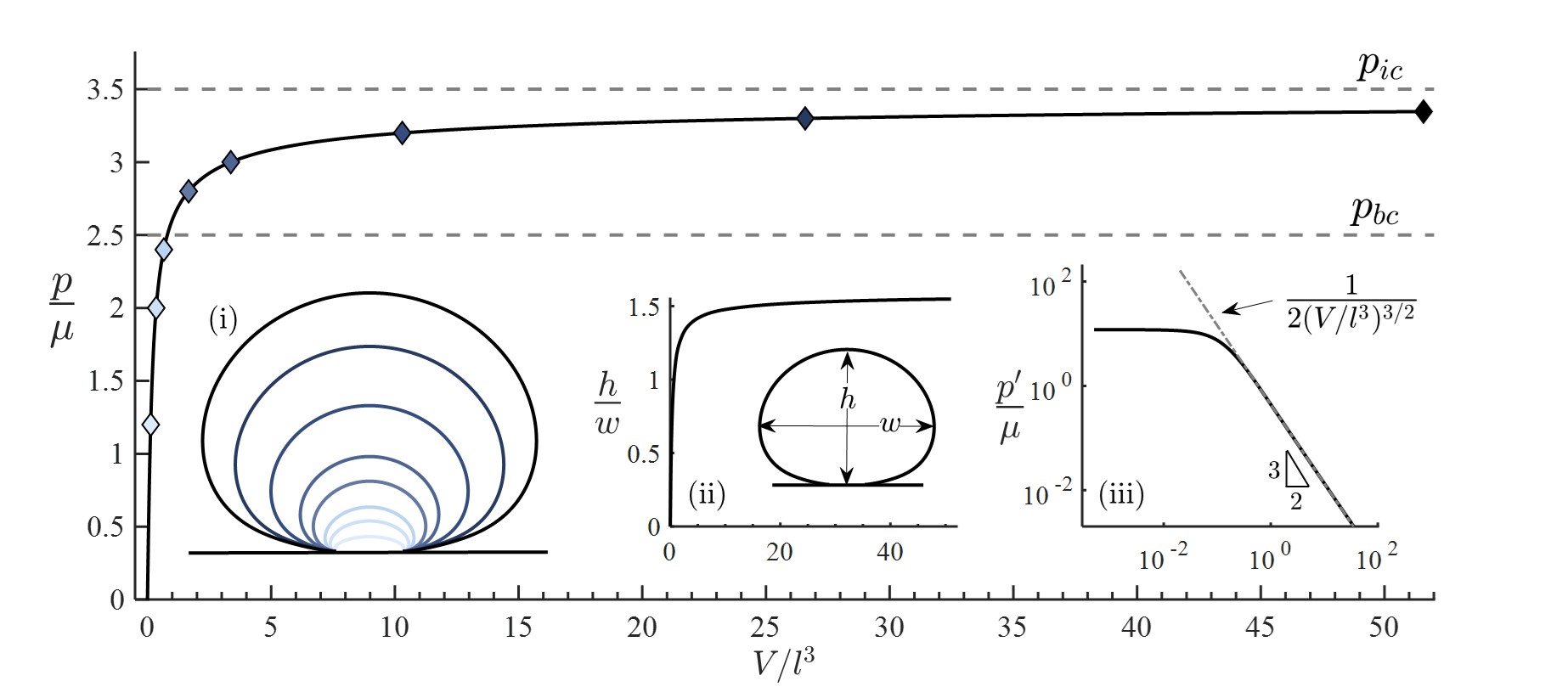}
\caption{
\textbf{The applied cavity pressure approaches an asymptotic limit of $p_{ic}/\mu\to7/2$ with increasing volumes.} The shape of the cavity at different dimensionless expansion volumes is shown in inset (i), where curve shades correspond to the diamond markers along the pressure-volume curve. The aspect ratio of these shapes approaches an asymptotic value of $h/w=1.6$, as shown in inset (ii). The power law decay of the slope of the pressure-volume curve, shown via the log-log plot in  inset (iii), confirms the asymptotic behavior. The data set for the pressure-volume curve and a video of the simulated expansion process can be found in \textcolor{blue}{\href{https://www.dropbox.com/s/io686v3hob031or/IC_Video.mp4?dl=0}{Supplementary Information}.}}
\label{fig1}
\end{figure*}


However, despite the vast body of literature on cavitation in solids that has accumulated over the years \cite{gent1959internal,ball1982discontinuous, chung1986finite, gent1990cavitation, huang1991cavitation, hang1992cavitation, horgan1995cavitation,barney2020cavitation} and in contrast to the study of cavitation in fluids; theoretical predictions in solids commonly consider cavitation that  initiates from defects in the bulk of an otherwise homogeneous material, while cavitation that initiates from interfaces are only rarely mentioned \cite{cohen2018competing,ringoot2021stick,wahdat2022pressurized}. Nonetheless,  defects are often found around impurities and at interfaces, and thus generating cavities far from interfaces can be challenging \cite{poulain2017damage}. Moreover, with the growing application of composites that integrate multiple materials  as well as layer-by-layer additive manufacturing methods, the use of materials that are prone to  failure at interfaces is becoming increasingly ubiquitous; in these materials initiation of failure from cavitation at interfaces may dominate over cavitation in the bulk. 
There are several possible explanations for the existing gap in providing a theory for interfacial cavitation that is as simple and elegant as cavitation in the bulk \cite{gent1959internal,ball1982discontinuous,huang1991cavitation,hang1992cavitation}. When a defect is embedded in the bulk of a material, its expansion can be well described using simplifying assumptions on the symmetry of the deformation field. This luxury must be forfeited when attempting to consider cavities at an interface, thus requiring a new approach to define the problem setting, and the aid of computational tools to capture the expansion process deep into the nonlinear range of the material response. Most importantly, it is not obvious that an asymptotic pressure even exists and if the lengthscale independent property of bulk cavitation  translates to interfacial cavitation. Finally, a major driver of earlier studies on cavitation has been their observation \cite{gent1959internal,andrews1973mechanics,ashby1989flow}. However, interfacial failure, even when it occurs locally, has not been previously considered from the viewpoint of cavitation, and has instead been interpreted as an interfacial fracture and delamination process \cite{suo1990singularities,hutchinson1991mixed,leung2001delamination,buyukozturk2004progress,chen2010approaches,buyukozturk2011structural}. 

To remedy this gap our work presents \textit{(i)} a complete theoretical framework that is analagous to bulk cavitation, thus exposing a lengthscale independent interfacial cavitation limit  and capturing the ensuing delamination process; \textit{(ii)} a phase diagram that determines the competition between bulk and interfacial cavitation and the stability thresholds across a broad range of normalised presssures and interface properties; and \textit{(iii)}   observations of interfacial cavitation and  delamination for a direct comparison and validation of the theory. 

To establish interfacial cavitation as a lengthscale independent process we exploit two experimental systems. At the small scale, with defects of the order of  $\sim 10\mu$m, we examine the growth of biofilm, starting from one \textit{Vibrio cholerae} bacterium embedded at the interface between a soft material (agarose) and a glass substrate to which it is bonded. In this highly controlled growth process, the evolving biofilm behaves like an expanding fluid as it proliferates \cite{li2022nonlinear}, allowing the tracking of morphological changes during growth\cite{zhang2021morphogenesis}. At larger scales of $\sim 100\mu$m we exploit the Pressurized Interfacial Failure (PIF) experimental set-up, recently described in \citep{wahdat2022pressurized}, whereby interfacial separation is controlled by applying fluid pressure at a localized region of a bonded interface.

\section*{An asymptotic interfacial cavitation pressure}
A key characteristic of the bulk cavitation pressure is that it is universal in the sense that it does not depend on the size of the defect. This lengthscale independence arises from the assumption that the defect is small compared to the size of the body, and neglects the effect of surface tension\footnote{It has been shown that if surface tension is present, it can resist cavitation, thus making larger defects more prone to the cavitation instability \cite{gent1959internal}.}. Using the same basic assumptions, in this work we consider a semi-infinite body that is bonded to a rigid substrate. We prescribe a defect as being a circular region of the interface, of diameter $l$, that is not bonded. To establish the existence of a critical interfacial cavitation pressure, we first assume that other regions of the interface are strictly bonded; the consequence of relaxing this assumption will be considered in the next section. Accordingly, we define our cartesian coordinate system $(X,Y,Z)$ with its origin at the center of the defect, such that the undeformed body occupies the region $X,Y\in(-\infty,\infty)$, $Z\in[0,\infty)$, and with the defect in the range  $X^2+Y^2\leq l^2/4$, $Z=0$. We assume that axial symmetry  is preserved as the interfacial cavity expands, {which is also supported by our experimental observations}.

The analysis presented in this paper can be conducted using any constitutive model of choice. Nonetheless, to compare our results with the well established neo-Hookean bulk cavitation pressure,  we limit our attention to incompressible neo-Hookean response. Thus we have the free energy function $\psi=\mu(I_1-3)/2$, where $I_1$ is the first invariant of the left Cauchy-Green deformation tensor.

As in the case for bulk cavitation, there are two ways to expand the cavity. By application of internal pressure at the cavity wall, or by application of remote tension. For bulk cavitation in incompressible materials, analytical derivations show that the cavitation pressure is the same for both scenarios\citep{knowles1965finite,cohen2015dynamic}. The nonlinear reciprocal theorem\cite{henzel2022reciprocal}, confirms that the same is true also for interfacial cavitation. Hence, in our finite element simulations we quasistatically expand the interfacial cavity by application of pressure at the  wall of the defect. This procedure is implemented in the Finite Element (FE) software ABAQUS/CAE 2017. To eliminate boundary effects, we set the size of the domain to 100 times the initial length of the defect and we have confirmed that changes in the remote field remain negligible. Quadratic, axisymmetric and hybrid elements (CAX8H) are used to capture large axisymmetric deformations and incompresibility. The mesh is highly refined  around the defect, such that the length of the smallest element is $l/10^{5}$.



As shown in Fig. \ref{fig1}, our simulation reveals that  the normalized interfacial cavity pressure $(p/\mu)$ approaches an asymptotic value of $p_{ic}/\mu\sim 7/2$ that is analogous to that obtained for  cavity expansion in the bulk. Our analysis confirms that this trend is indeed asymptotically approaching a finite value. In addition to refining our computations to access pressures in the range of extreme deformations with a volume expansion $V$ as large as $50l^3$,  we consider the evolution of its slope  $p'={\rm d}p/{\rm d} (V/l^3)$ on a log-log plot and find that it decays like $(V/l^3)^{-3/2}$. Assuming this asymptotic decay,  for large dimensionless volumes $V/l^3\gg1$, the evolution of pressure takes the simple form $p\cong p_{ic}-\mu/\sqrt{V/l^3}$. 

In examining the evolving shape of the cavity as it expands, we find that, in contrast to bulk cavitation\cite{durban1997spherical}, a self-similar field does not appear to emerge, even deep into the nonlinear range. Nonetheless, the aspect ratio of the cavity saturates at a value of $\sim 1.6$,  and does not tend towards a spherical shape. This limiting value of the aspect ratio can be used to determine whether an experimental observation is approaching the cavitation limit and to thus infer whether the pressure within the cavity is approaching $p_{ic}$.
A video of the expansion process is provided in the \textcolor{blue}{\href{https://www.dropbox.com/s/io686v3hob031or/IC_Video.mp4?dl=0}{Supplementary Information},} where the stress concentration is shown by the darker shades, which represent variation in the  second invariant of the deviatoric stress tensor field.  

We anticipate that by considering a rigid substrate, the interfacial cavitation pressure obtained here serves as an upper bound for the case of deformable substrates, whereas the case of a substrate that has stiffness identical to that of the body, will cavitate at a pressure that is well represented by the spherical cavitation limit. This is confirmed by earlier studies that show that the initial shape of the defect has little influence on the cavitation pressure\cite{raayai2019volume}. Nonetheless, additional work is needed to elucidate the relationship between the stiffness ratio of the two materials, and the resulting interfacial cavitation pressure, which is beyond the scope of this work.

It should be noted that the extreme deformations that are achieved in the vicinity of the cavity, as  the asymptotic value of the cavitation pressure is approached, are unlikely to be well represented by a neo-Hookean model. This is true also in the case of bulk cavitation. Nonetheless, the internal pressure is determined by the resistance of the entire field, and  thus the neo-Hookean model is considered a good approximation, even if some inelastic effects occur in a localized region near the cavity wall. 
A competition between different instability modes may emerge, be it cavitation in the bulk, at the interface, or alternatively failure of the interfacial bond, as discussed next\footnote{In this work we do not consider the possible onset of fracture in the bulk of the material, which has been the focus of an earlier study \cite{raayai2019intimate}.}.

\begin{SCfigure*}[\sidecaptionrelwidth][t]
\centering
\includegraphics[width=1.3\linewidth]{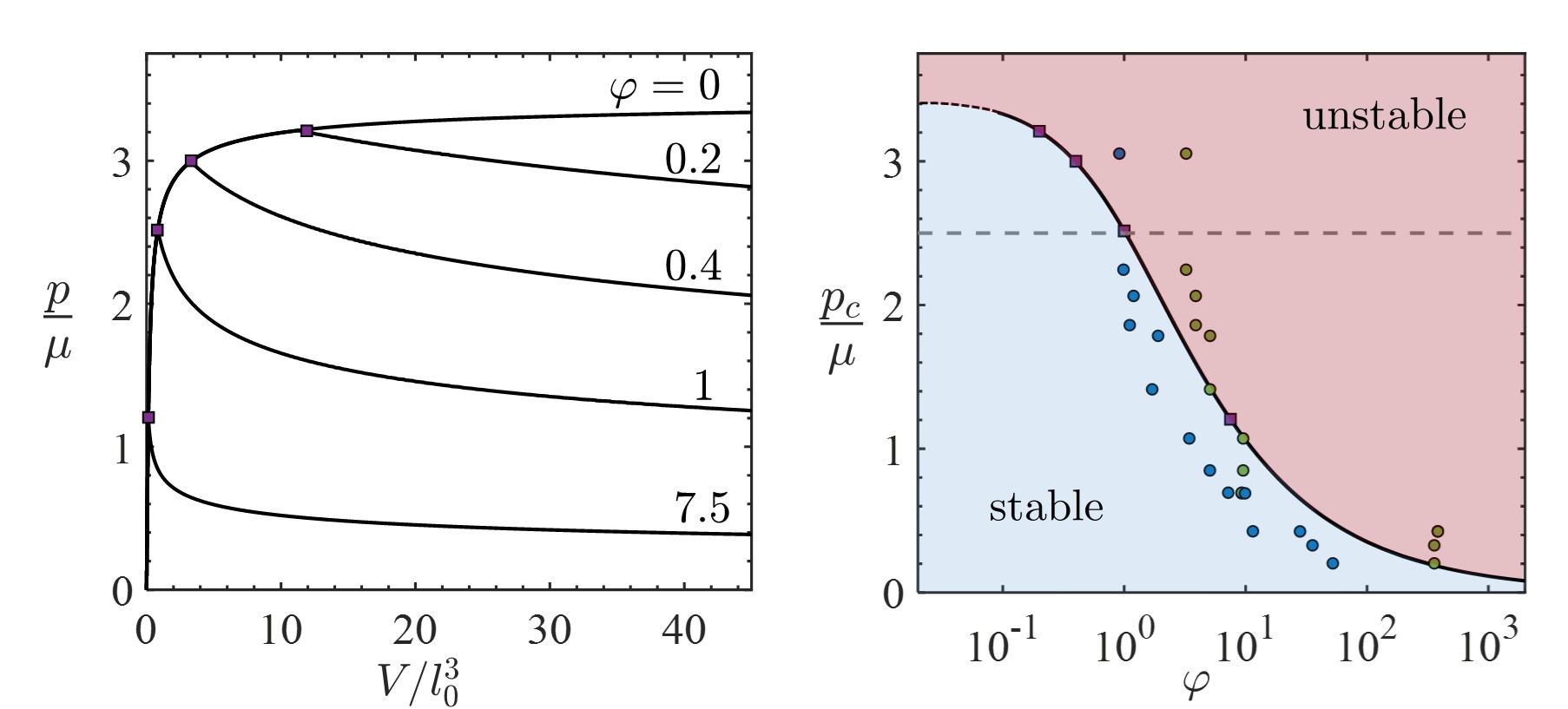}
\vspace{-3mm}\caption{\textbf{The dimensionless material property $\varphi=\mu l_0/\Gamma$ determines the stability threshold.} On the left, departure of the pressure a from the purely elastic response is shown for various values of $\varphi$, and indicted by the square markers. The corresponding critical pressure $(p_c/\mu)$ is shown as a function of $\varphi$ on the right, to form a phase diagram, with the square markers corresponding to the curves on the left. In the blue region the response  is stable; in the red region  perturbations can lead to unstable expansion. The dashed line corresponds to the bulk cavitation pressure.  The circular markers represent critical pressures measured in different materials using the PIF method, where the interface toughness is determined via  conventional
probe-tack test (blue) or linear theory\cite{wahdat2022pressurized} (green).  Experimental details can be found in the Supplementary Information.
}

\label{fig:side}
\label{fig2}
\end{SCfigure*}

\section*{Unstable cavity growth }
The presence of an asymptotic value of the interfacial pressure, as shown in Fig. \ref{fig1}, implies that upon approaching $p_{ic}$, small perturbations of pressure can induce substantial changes in cavity volume, namely  unstable growth. By comparing  interfacial cavitation  with  bulk cavitation, we find that locally pressurized interfacial cavitities can withstand pressures that would `rip the material apart' if applied in the bulk. This striking  $40 \%$ increase in normalized critical pressure (i.e. $p_{ic}\cong1.4p_{bc}$) can be attributed to the additional strength that is provided by the rigid constraint of the substrate. However, if the strength of the interfacial bond is finite, failure of the interface may initiate first. 

To examine the possible onset of interfacial failure, we consider the energetically favorable states of the system for a given prescribed volume. The system is characterised by the two independent state variables: the dimensionless volume ($V/l^3$) and the defect length ($l$). Central to our formulation is the fact that the elastic expansion process is length scale insensitive for the assumption of semi-infinite body and that the relationship between the dimensionless pressure and the dimensionless volume, plotted in Fig. \ref{fig1}, applies for cavities of any defect length. 
Accordingly, we write the relationship between the dimensionless pressure and the dimensionless volume in Fig. \ref{fig1} as $p/\mu=f'(V/l^3)$, where the prime denotes differentiation. The total elastic energy in the system can thus be directly written as $E_e=\int_0^V p{\rm d} V=\mu l^3 f(V/l^3)$. 



If delamination is permitted, an additional energetic cost would be incurred to create new surface area, which we write as $E_d=\Gamma\frac{\pi}{4}(l^2-l_0^2)$, where $l_0$ denotes the initial defect diameter, and $\Gamma$ is the energy needed to debond a unit area of the surface -- the interface toughness. The total energy invested in expanding the cavity is thus the sum of the two contributions $E_t=E_e+E_d$, which can be written in its dimensionless form $\mathcal{E}_t=E_t/(\Gamma l_0^2)$ as \vspace{-3mm}
\begin{equation}\label{enrg}
     \mathcal{E}_t(l;l_0,V)= \varphi \left(\frac{l^3}{l_0^3}\right) f(V/l^3)+ \frac{\pi}{4}  \left(\frac{l^2}{l_0^2}-1\right)\vspace{-2mm}
\end{equation}
 where the dimensionless model parameter $\varphi=\mu l_0/\Gamma$ determines the balance between elastic and interfacial effects. 
For a prescribed volume $V$, the energy in \eqref{enrg} depends only on the deliminated size of the defect $l$, which in an equilibrium configuration will minimize the total energy\footnote{Note that this this energetic argument only holds for the forward process of delamination. Upon unloading, additional considerations may apply in describing the re-engagement of the two surfaces \cite{ringoot2021stick}.}. Formally, this can be written as \vspace{-2mm}\begin{equation} l={\rm arg}~  \underset{l}{\rm min} \left(\mathcal{E}_t(l;l_0,V,\varphi) \right),\vspace{-2mm}\end{equation}
which can be evaluated by taking the derivative, such that  ${\partial}\mathcal{E}_t/{\partial}l=\varepsilon_t(l;l_0,V,\varphi)=0$, and the corresponding pressure is  obtained from the pressure volume curve (Fig. \ref{fig1}). 
By analyzing this functional form, we find that the onset of interfacial failure is a second order transition. We can thus evaluate the critical conditions for initiation of failure by substituting $l=l_0$, to write $\varepsilon_t(l_0;l_0,V,\varphi)=0$, which provides us with an implicit relationship between the normalized critical volume and the model parameter, $V_c/l^3_0=g(\varphi)$. The corresponding critical pressure is then obtained as $p_c(\varphi)/\mu=f'(g(\varphi))$.

Fig. \ref{fig2} shows the evolution of pressure with increasing cavity volume for various values of $\varphi$. The curves  deviate from the response for purely elastic expansion ($\varphi=0$) at the  critical pressure  $p_c$,  as marked by purple squares.  The decrease in pressure along the equilibrium branch is explained by the fact that although the volume is increasing, the normalized volume $(V/l^3)$ decreases, and thus motion along the pressure-volume curve (Fig. \ref{fig1}) changes direction.  
It is important to note that  before onset of delamination the response  is independent of whether the loading is applied via pressure control or volume control and adheres to the pressure-volume curve. At onset of delamination, a pressure controlled system would exhibit indefinite unstable expansion for any $p>p_c$, whereas a volume controlled system would proceed along the descending branch.

The stable and unstable regimes of the system are shown on a phase diagram  (Fig. \ref{fig2}). With the black curve indicating the critical pressure $p_c(\varphi)/\mu$; the blue and red shades distinguish stable and unstable regions, respectively. For strictly bonded interfaces $(\varphi\to 0)$, we recover the interfacial cavitatation pressure; for vanishingly weak bonding, interfacial failure would dominate ($\varphi\to\infty$). At the latter limit,  the critical pressure scales as $1/\sqrt{\varphi}$. This result agrees  with  linear fracture models that are commonly used to model delamination phenomena \cite{suo1990singularities,hutchinson1991mixed,leung2001delamination,buyukozturk2004progress,chen2010approaches,buyukozturk2011structural}. Markers on the phase diagram show experimental results from PIF experiments, as described next. 






\section*{Interfacial cavitation explains PIF experiments}
The competition between bulk and interfacial contributions in interfacial separation processes, and the nonlinear effects that ensue, have long been a hindrance on the development of methods for characterization of adhesion properties, where decoupling of the different effects is essential\cite{lakrout1999direct,zosel1989adhesive}.  
The Pressurized Interfacial Failure (PIF) method was recently proposed to combat this issue\citep{wahdat2022pressurized}.
In the PIF experimental set-up, soft adhesives are compressed and then locally pressurised in a small region of an otherwise bonded interface. As pressure  increases the growth of an interfacial cavity is observed and appears as a circular delamination front, thus preserving axisymmetry, as assumed in our model. By identifying the pressure at initiation of instability $(p_c)$ and tracking the front propagation, this measurement technique pairs with a model that assumes a linear elastic response of the material to determine the interfacial toughness. PIF data is obtained for two materials systems: acrylic elastomers composed of poly(n-butyl)acrylate (PBA) networks cross-linked by
ethylene glycol dimethacrylate (EGDMA), and commercial VHB tape,  to explore a range of bulk and interfacial properties (see Supplementary Information for details). The circular markers on the phase diagram represent experimental measurements of the critical pressure measured using this technique (Fig. \ref{fig2}). The corresponding material parameter $(\varphi)$ is determined using a conventional probe-tack test (blue markers) or the PIF technique (green markers). 

A remarkable agreement between the experimental measurements and the theoretical curve is observed. The theory serves as an upper-bound on the probe-tack data, whereas the PIF method appears to overestimate $\phi$ by underestimating the interface toughness. Quite interestingly, for higher interface toughness (lower $\varphi$), the experiments exhibit high values of critical pressure that exceed the bulk cavitation limit and approach the interfacial cavitation limit. This is a clear indication of highly nonlinear deformations that cannot be captured  by linear models. Moreover, this demonstrates that interfacial cavitation, although a theoretical limit, is approached in synthetic material systems. Next, we will show that interfacial cavitation may be  crucial also in determining the fate of natural systems, as we focus our attention to interfacial growth of bacterial biofilms.

\section*{Cavitation to delamination transition observed in biofilm}
Bacteria are often found on interfaces: at the site of a wound\cite{kirketerp2008distribution,gjodsbol2006multiple}; on a surgical implant\cite{buret1991vivo}; and in various chronic infections \cite{costerton1999bacterial,rybtke2015pseudomonas}.
One  cell embedded on an interface can multiply  and form a biofilm consisting of tens of thousands of bacteria. This process inevitably deforms the surrounding medium, pushing it away from the substrate, and making room for reproduction. With most observations of biofilm growth conducted in absence of such resistance, on flat substrates\cite{garrett2008bacterial, seminara2012osmotic, fei2020nonuniform, song2015effects}, the confined growth of biofilm in three dimensional settings has only been observed in recent years \cite{Zhange2107107118} and has focused on the growth of biofilms as inclusions that are embedded in the bulk of a medium \cite{li2022nonlinear}. However, less is known about the confined growth of biofilm  at an interface; a situation which could be a better representation of \textit{in vivo} conditions \cite{bjarnsholt2013vivo}. Our theory suggests that such growth could lead to mechanical instability; the particular mode of instability, and whether or not it is achieved,  can influence the growth path of the biofilm and therefore alter its developmental trajectory. 

\begin{figure}[h!]\hspace{-4mm}
\includegraphics[width=1.15\linewidth]{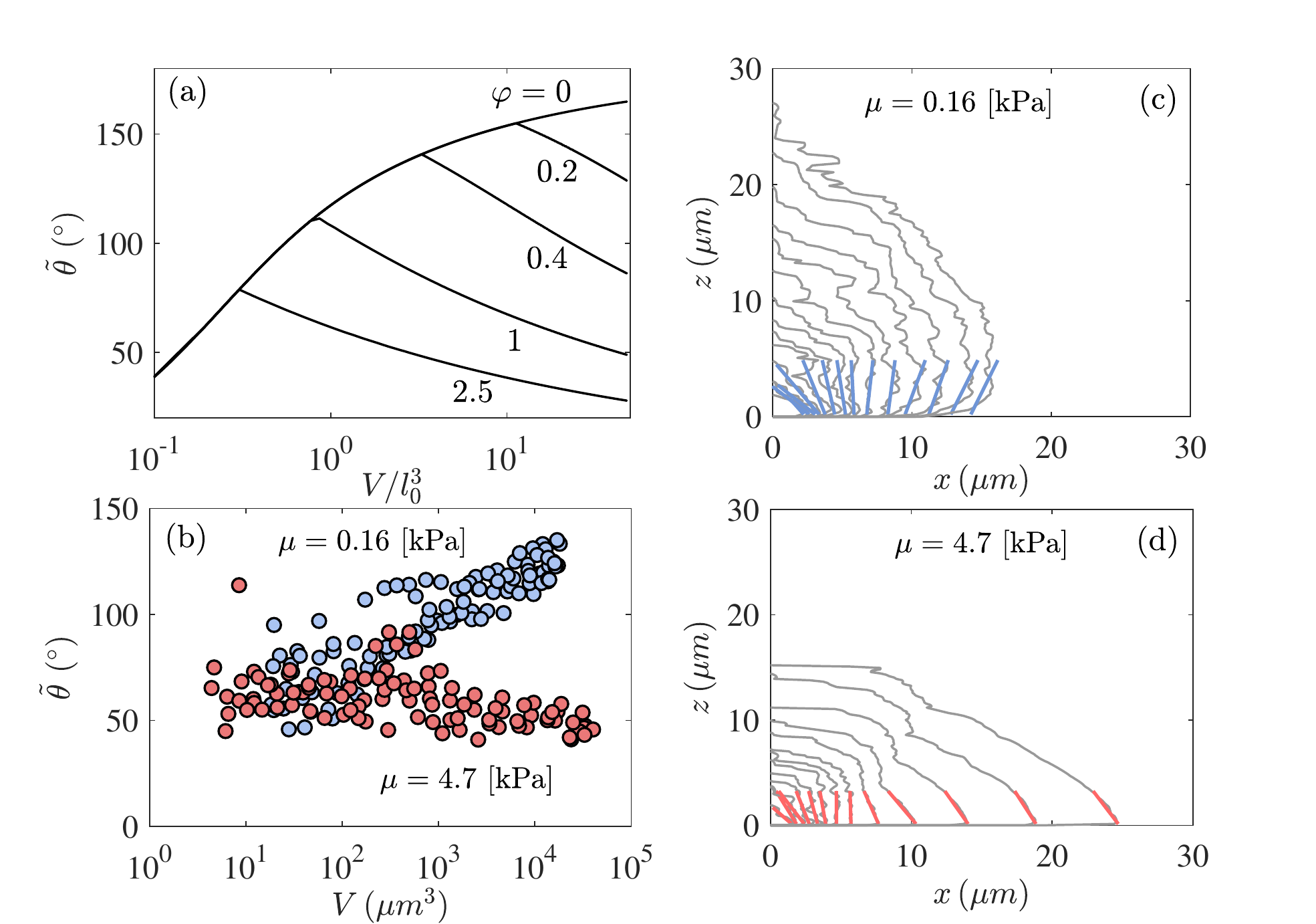}
\caption{\textbf{Evolution of interstitial biofilm exhibits a cavitation to delamination transition.} (a) Theoretically predicted evolution of apparent angle, for increasing volume, shown for varying values of $\varphi$. (b) Experimentally measured apparent angle of biofilm of different volumes shown for two stiffnesses, $\mu=0.16$ and $4.7$ [kPa], as marked by blue and red markers, respectively. (c,d) Shape evolution of a single biofilm, for each of the two stiffnesses. The colored lines represent the experimentally estimated apparent angles.
}
\label{fig3}
\end{figure}

Beyond its potential medical implications, observing the growth of biofilm at an interface also provides a unique opportunity to examine interfacial cavitation at the small scales. 
To this end, in our experimental system, isolated \textit{Vibrio cholerae} bacteria are embedded at the interface between a glass substrate and soft agarose gels of varying stiffness. Starting from one cell, we use an adaptive imaging technique\cite{yan2018bacterial} that enables visualization of the global morphology over several orders of magnitude in volume (see details in Supplementary Information). 

\begin{figure}
\includegraphics[width=1\linewidth]{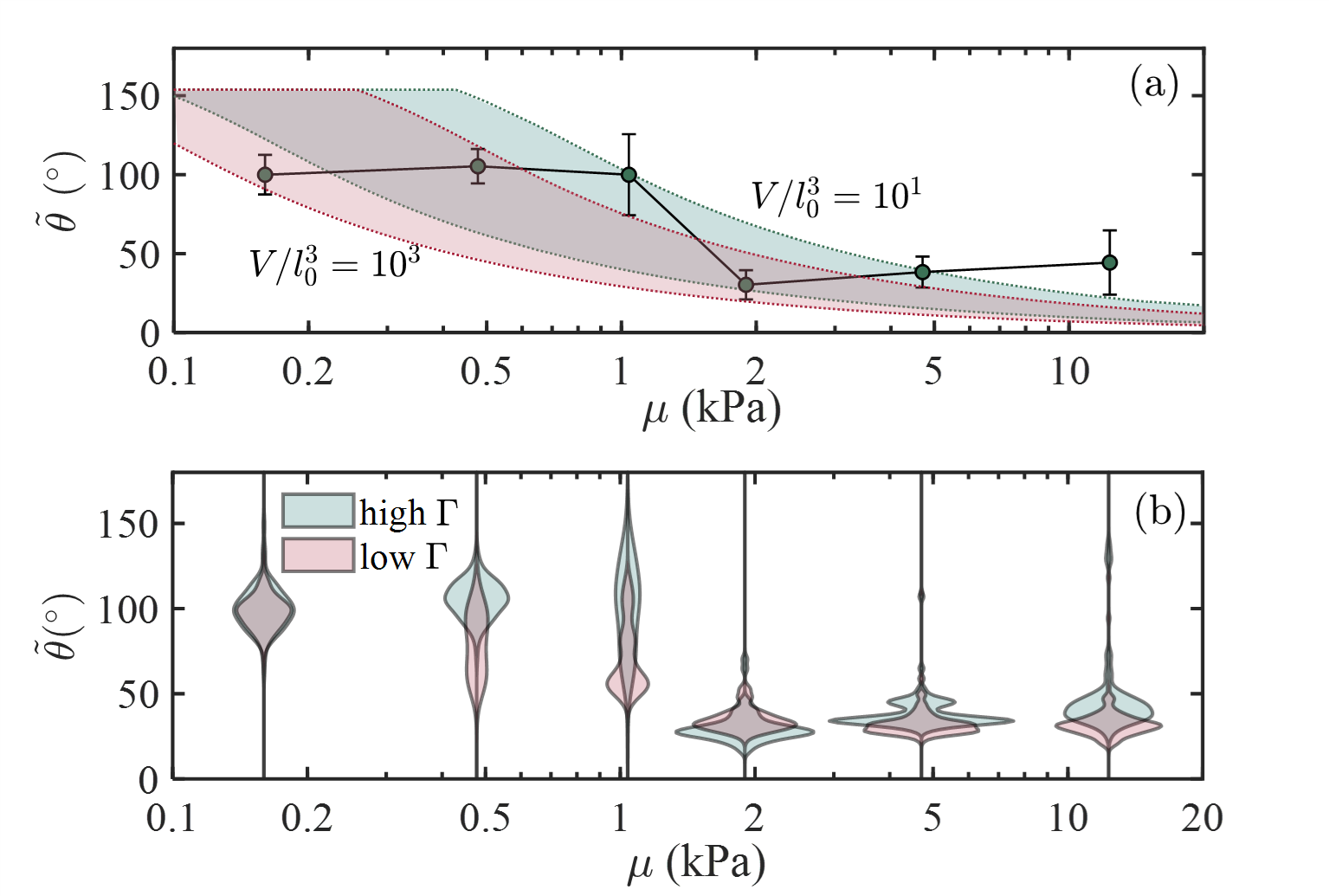}
\caption{\textbf{Influence of the interface toughness.} (a) Theoretical prediction of apparent contact angle is shown for a range of mature volumes (as indicated by the shaded regions) for biofilm grown in confinements of different stiffness.  Dashed lines correspond to two different volume expansions, $V=(10^4, 10^6)\mu$m$^3$, with red and green corresponding to different interface toughness, $\Gamma=(1.2,2)\times10^{-2}$N/m, respectively,  and using $l_0=10\mu$m as the initial defect size. For comparison, the green markers represent the experimentally measured value for the system with the higher $\Gamma$, and error bars show the standard deviations.  (b) Experimentally measured mature apparent angles  for two systems of different interface toughness are shown in the form of a violin plot, whereby the width of the cords are reflective of the probability density. 
}
\label{fig4}
\end{figure}

While biofilm have measurable solid-like mechanical properties\cite{yan2018bacterial}, a recent study\cite{li2022nonlinear} has shown that on the timescale of growth, the persistent internal reorganization of cells in response to the mechanical confinement imparts the biofilm with fluid-like properties that dictate the evolution of its macroscopic shape. This implies that the reaction force between the biofilm and the confinement can be modeled as a hydrostatic pressure, as considered herein. Using this insight, we compare the theoretically predicted evolution of the interfacial cavity shapes with those observed in the biofilm system (Figs. \ref{fig3},\ref{fig4}). To quantify the shapes we focus our attention to the \textit{apparent contact angle}, which we define as the internal angle at the intercept of the cavity contour with the interface. Since in the simulation the local angle can be highly influenced by discretization, for consistency we determine it as the uniquely defined  intercept angle  of a spheroid that intersects the interface at the same location and crosses through the peak location at $X=0$.  
Accordingly, we show the predicted evolution of the apparent contact angle, with increasing volume (log scale) in Fig. \ref{fig3}(a). If no delamination occurs the angle monotonically increases. At onset of delamination a clear transition to a decreasing trend is observed, and occurs earlier for increasing values of $\varphi$.

The theoretical trends are mirrored by our experimental observations of biofilm growth, as shown by comparing  Fig. \ref{fig3}(a) with the experimental curves in Fig. \ref{fig3}(b). For lower stiffness of the confining gel (blue markers), the apparent angle continues to increase with increasing volume, thus approaching the interfacial cavitation limit. For the stiffer gel (red markers), the initial trend at small volumes resembles that of the softer gel, until a clear departure is observed (at $V\sim 300~\mu$m with $\tilde{\theta}\sim75^{\circ}$). The following  monotonic decay indicates  progression of delamination. 
A typical shape evolution of one biofilm for each of the two  stiffnesses is shown in Fig. \ref{fig3}(c,d). The colored lines represent the apparent contact angle, which for the experimental curves, is estimated via a local linear fit.

Comparing the location of the transition point in Fig. \ref{fig3}(b) with the theorectical curves in Fig. \ref{fig3}(a) allows us to obtain an approximate measure of the initial defect size induced by the seed bacterium. Onset of delamination at  $\tilde{\theta}\sim75^{\circ}$ corresponds to the curve with $\varphi\sim 2.5$, for which the  transition occurs at the dimensionless volume $V/l_0^3 \approx 0.3$. Since the dimensional volume at the transition  is $V\sim 300~\mu \text{m}^3$, we have that $l_0\sim 10 \mu$m, corresponding to a small biofilm with tens of cells. Next, since the agarose stiffness is known ($\mu=4.7[kPa]$), along with $\varphi$ and $l_0$, we can estimate the interfacial toughness as $\Gamma\sim2\times10^{-2}$N/m.  Note that the effective initial defect size $(l_0)$ is reflective of the length at which the biofilm is established and can be treated as a continuum body. In our experiments, we track the shapes of the biofilm starting from a single bacterium, as seen in Fig. \ref{fig3}(c,d).

Finally, we examine the influence of the interfacial toughness on  biofilm shapes in Fig. \ref{fig4}. The apparent contact angles of hundreds of mature biofilms, grown under agarose gels of varying stiffnesses, are recorded for the system in Fig. \ref{fig3}, and also for a system in which we apply a treatment to the glass surface, to reduce the interface toughness (see Supplementary Information for details). We define a mature biofilm as one that has grown for $12-16$ hours. After this growth period we find that  biofilm volumes  can vary in the range $V\in(10^4,10^6)\mu$m$^3$. Assuming that the initial defect size remains of the same order (i.e. $l_0\sim10\mu$m), we show the corresponding theoretically predicted range of apparent contact angles for two values of interface toughness, $\Gamma=(1.2,2)\times10^{-2}$N/m, in Fig. \ref{fig4}(a). The latter corresponds to the results in Fig. \ref{fig3} and is shaded green. The former represents a slight reduction in toughness, and is shaded red.  The experimental results are shown in the form of a \textit{violin plot} in Fig. \ref{fig4}(b), where the probability density of different apparent angles is represented by the width of the respective cords. For visual comparison, the average apparent angles for the system with higher interface toughness are also shown in Fig. \ref{fig4}(a), where error bars represent standard deviations\footnote{Note that due to existing limitations in measurement techniques, we were not able to obtain an independent experimental measure of the $\Gamma$ values, with sufficient resolution.}.  

Both the theoretical and the experimental results reveal  that in softer environments, mature biofilms maintain large contact angles. The initial plateau at small stiffnesses suggests that the gel-substrate interface remains mostly intact.  A clear transition to smaller angles is observed for the stiffer confinements. We attribute the more pronounced transition in the experimental data to the unstable nature of the delamination process, which promotes diffusion of solvent into the cavity, thus relaxing the volume constraint. 
By comparing results for the different interface toughness, both the theory and the experiments   exhibit an earlier and smoother transition to the flat, delaminated, shapes for the system with lower interface toughness.

Overall, we find that biofilms may deform the confining medium significantly, thus approaching the interfacial cavitation limit, before  delamination is triggered. Both the material stiffness and the interface toughness play important roles in determining the fate these microscopic growing entities. 

\section*{Conclusions}
In solids, cavitation that occurs at interfaces has received far less attention than bulk cavitation, which has become a well established criteria for onset of failure in various materials, ranging from ductile metals to biological tissue. Though observations of interfacial failure have been reported, and can become dominant in heterogeneous materials, it has been primarily studied as a process of adhesive failure, which is length scale dependent. Our theoretical model shows that, in analogy to bulk cavitation, interfacial cavity expansion also arrives at a scale free asymptotic limit -- the interfacial cavitation pressure -- $p_{ic}/\mu=7/2$, beyond which the expansion becomes unstable. Observations in two different experimental systems confirm that this theoretical limit is indeed approached, thus supporting the need for  theory that can predict the highly nonlinear material response. Beyond cavitation, our theory extends to understand the propagation of delamination,  thus providing a phase diagram that is useful to distinguish between stable and unstable regimes, and compares  well with the experimental measurements. Given that interfacial failure is a ubiquitous phenomenon relevant to both synthetic and natural systems, we anticipate that by identifying the stability limits on their load bearing capacity will enable informed interpretation of experimental observations and can pave the way  for  design of more resilient heterogeneous material systems. Finally, this work is not without limitations, there is a breadth of opportunities to extend this work to examine the response for different constitutive models  that incorporate strain stiffening, inelastic deformation, or compressibility, as well as extending the framework  to capture the role of non-rigid substrates or inertia.



\acknow{T.H. and T.C. acknowledge the support of Dr. Timothy B. Bentley, Office of Naval
Research Program Manager, under award number N00014-20-1-2561, and support from the National Science Foundation under
award number CMMI-1942016. C.S. and T.C. acknowledge the support from the National Science Foundation (CMMI, MOMS, 1942016). J.Y. holds a Career Award at the Scientific Interface from the Burroughs Welcome Fund, and acknowledges the support of the National Institute of General Medical Sciences at the National Institutes of Health, under Award Number DP2GM146253. H.W. and A.J.C acknowledge the financial support by Saint-Gobain through the Center for UMass/Industry Research on Polymers (CUMIRP).}

\showacknow{} 



\section*{References}
\begin{thebibliography}{10}

\bibitem{rayleigh1917viii}
Rayleigh L (1917) Viii. on the pressure developed in a liquid during the
  collapse of a spherical cavity.
\newblock {\em The London, Edinburgh, and Dublin Philosophical Magazine and
  Journal of Science} 34(200):94--98.

\bibitem{besant1859treatise}
Besant WH (1859) {\em A treatise on hydrostatics and hydrodynamics}.
\newblock (Deighton, Bell).

\bibitem{bishop1945theory}
Bishop R, Hill R, Mott N (1945) The theory of indentation and hardness tests.
\newblock {\em Proceedings of the Physical Society (1926-1948)} 57(3):147.

\bibitem{puttick1959ductile}
Puttick K (1959) Ductile fracture in metals.
\newblock {\em Philosophical magazine} 4(44):964--969.

\bibitem{ashby1966work}
Ashby M (1966) Work hardening of dispersion-hardened crystals.
\newblock {\em Philosophical Magazine} 14(132):1157--1178.

\bibitem{tanaka1970cavity}
Tanaka K, Mori T, Nakamura T (1970) Cavity formation at the interface of a
  spherical inclusion in a plastically deformed matrix.
\newblock {\em The Philosophical Magazine: A Journal of Theoretical
  Experimental and Applied Physics} 21(170):267--279.

\bibitem{le1981model}
Le~Roy G, Embury J, Edwards G, Ashby M (1981) A model of ductile fracture based
  on the nucleation and growth of voids.
\newblock {\em Acta Metallurgica} 29(8):1509--1522.

\bibitem{tvergaard1982ductile}
Tvergaard V (1982) Ductile fracture by cavity nucleation between larger voids.
\newblock {\em Journal of the Mechanics and Physics of Solids} 30(4):265--286.

\bibitem{kassner2003creep}
Kassner M, Hayes T (2003) Creep cavitation in metals.
\newblock {\em International Journal of Plasticity} 19(10):1715--1748.

\bibitem{murali2011atomic}
Murali P, et~al. (2011) Atomic scale fluctuations govern brittle fracture and
  cavitation behavior in metallic glasses.
\newblock {\em Physical Review Letters} 107(21):215501.

\bibitem{shen2021observation}
Shen LQ, et~al. (2021) Observation of cavitation governing fracture in glasses.
\newblock {\em Science Advances} 7(14):eabf7293.

\bibitem{chen2002deep}
Chen X, Li Q (2002) Deep penetration of a non-deformable projectile with
  different geometrical characteristics.
\newblock {\em International Journal of Impact Engineering} 27(6):619--637.

\bibitem{masri2005dynamic}
Masri R, Durban D (2005) Dynamic spherical cavity expansion in an elastoplastic
  compressible mises solid.
\newblock {\em Journal of Applied Mechanics} 72(6):887--898.

\bibitem{forrestal2008penetration}
Forrestal MJ, Warren TL (2008) Penetration equations for ogive-nose rods into
  aluminum targets.
\newblock {\em International Journal of Impact Engineering} 35(8):727--730.

\bibitem{borvik2009perforation}
B{\o}rvik T, Forrestal M, Hopperstad O, Warren T, Langseth M (2009) Perforation
  of aa5083-h116 aluminium plates with conical-nose steel
  projectiles--calculations.
\newblock {\em International Journal of Impact Engineering} 36(3):426--437.

\bibitem{cohen2010ballistic}
Cohen T, Masri R, Durban D (2010) Ballistic limit predictions with quasi-static
  cavitation fields.
\newblock {\em International Journal of Protective Structures} 1(2):235--255.

\bibitem{crosby2011blowing}
Crosby AJ, McManus J (2011) Blowing bubbles to study living material.
\newblock {\em Physics Today} 64(2):62--63.

\bibitem{zimberlin2007cavitation}
Zimberlin JA, Sanabria-DeLong N, Tew GN, Crosby AJ (2007) Cavitation rheology
  for soft materials.
\newblock {\em Soft Matter} 3(6):763--767.

\bibitem{zimberlin2010cavitation}
Zimberlin JA, McManus JJ, Crosby AJ (2010) Cavitation rheology of the vitreous:
  mechanical properties of biological tissue.
\newblock {\em Soft Matter} 6(15):3632--3635.

\bibitem{raayai2019volume}
Raayai-Ardakani S, Chen Z, Earl DR, Cohen T (2019) Volume-controlled cavity
  expansion for probing of local elastic properties in soft materials.
\newblock {\em Soft matter} 15(3):381--392.

\bibitem{chockalingam2021probing}
Chockalingam S, Roth C, Henzel T, Cohen T (2021) Probing local nonlinear
  viscoelastic properties in soft materials.
\newblock {\em Journal of the Mechanics and Physics of Solids} 146:104172.

\bibitem{gent1990cavitation}
Gent A (1990) Cavitation in rubber: a cautionary tale.
\newblock {\em Rubber Chemistry and Technology} 63(3):49--53.

\bibitem{lefevre2015cavitation}
Lef{\`e}vre V, Ravi-Chandar K, Lopez-Pamies O (2015) Cavitation in rubber: an
  elastic instability or a fracture phenomenon?
\newblock {\em International Journal of Fracture} 192(1):1--23.

\bibitem{poulain2017damage}
Poulain X, Lefevre V, Lopez-Pamies O, Ravi-Chandar K (2017) Damage in
  elastomers: nucleation and growth of cavities, micro-cracks, and
  macro-cracks.
\newblock {\em International Journal of Fracture} 205(1):1--21.

\bibitem{raayai2019intimate}
Raayai-Ardakani S, Earl DR, Cohen T (2019) The intimate relationship between
  cavitation and fracture.
\newblock {\em Soft matter} 15(25):4999--5005.

\bibitem{goriely2010elastic}
Goriely A, Moulton DE, Vandiver R (2010) Elastic cavitation, tube hollowing,
  and differential growth in plants and biological tissues.
\newblock {\em EPL (Europhysics Letters)} 91(1):18001.

\bibitem{kothari2020effect}
Kothari M, Cohen T (2020) Effect of elasticity on phase separation in
  heterogeneous systems.
\newblock {\em Journal of the Mechanics and Physics of Solids} 145:104153.

\bibitem{vidal2021cavitation}
Vidal-Henriquez E, Zwicker D (2021) Cavitation controls droplet sizes in
  elastic media.
\newblock {\em Proceedings of the National Academy of Sciences} 118(40).

\bibitem{kothari2022crucial}
Kothari M, Cohen T (2022) The crucial role of elasticity in regulating
  liquid-liquid phase separation in cells.
\newblock {\em arXiv preprint arXiv:2201.04105}.

\bibitem{gent1959internal}
Gent A, Lindley P (1959) Internal rupture of bonded rubber cylinders in
  tension.
\newblock {\em Proceedings of the Royal Society of London. Series A.
  Mathematical and Physical Sciences} 249(1257):195--205.

\bibitem{ball1982discontinuous}
Ball JM (1982) Discontinuous equilibrium solutions and cavitation in nonlinear
  elasticity.
\newblock {\em Philosophical Transactions of the Royal Society of London.
  Series A, Mathematical and Physical Sciences} 306(1496):557--611.

\bibitem{chung1986finite}
Chung DT, Horgan C, Abeyaratne R (1986) The finite deformation of internally
  pressurized hollow cylinders and spheres for a class of compressible elastic
  materials.
\newblock {\em International Journal of Solids and Structures}
  22(12):1557--1570.

\bibitem{huang1991cavitation}
Huang Y, Hutchinson J, Tvergaard V (1991) Cavitation instabilities in
  elastic-plastic solids.
\newblock {\em Journal of the Mechanics and Physics of Solids} 39(2):223--241.

\bibitem{hang1992cavitation}
Hang-Sheng H, Abeyaratne R (1992) Cavitation in elastic and elastic-plastic
  solids.
\newblock {\em Journal of the Mechanics and Physics of Solids} 40(3):571--592.

\bibitem{horgan1995cavitation}
Horgan CO, Polignone DA (1995) Cavitation in nonlinearly elastic solids: a
  review.
\newblock {\em Applied Mechanics Reviews} 48(8):471--485.

\bibitem{barney2020cavitation}
Barney CW, et~al. (2020) Cavitation in soft matter.
\newblock {\em Proceedings of the National Academy of Sciences}
  117(17):9157--9165.

\bibitem{cohen2018competing}
Cohen T, Chan CU, Mahadevan L (2018) Competing failure modes in finite adhesive
  pads.
\newblock {\em Soft matter} 14(10):1771--1779.

\bibitem{ringoot2021stick}
Ringoot E, Roch T, Molinari JF, Massart TJ, Cohen T (2021) Stick--slip
  phenomena and schallamach waves captured using reversible cohesive elements.
\newblock {\em Journal of the Mechanics and Physics of Solids} 155:104528.

\bibitem{wahdat2022pressurized}
Wahdat H, Zhang C, Chan N, Crosby AJ (2022) Pressurized interfacial failure of
  soft adhesives.
\newblock {\em Soft Matter}.

\bibitem{andrews1973mechanics}
Andrews E, Kinloch AJ (1973) Mechanics of adhesive failure. ii.
\newblock {\em Proceedings of the Royal Society of London. A. Mathematical and
  Physical Sciences} 332(1590):401--414.

\bibitem{ashby1989flow}
Ashby MF, Blunt FJ, Bannister M (1989) Flow characteristics of highly
  constrained metal wires.
\newblock {\em Acta Metallurgica} 37(7):1847--1857.

\bibitem{suo1990singularities}
Suo Z (1990) Singularities, interfaces and cracks in dissimilar anisotropic
  media.
\newblock {\em Proceedings of the Royal Society of London. A. Mathematical and
  Physical Sciences} 427(1873):331--358.

\bibitem{hutchinson1991mixed}
Hutchinson JW, Suo Z (1991) Mixed mode cracking in layered materials in {\em
  Advances in applied mechanics}.
\newblock (Elsevier) Vol.{}~29, pp. 63--191.

\bibitem{leung2001delamination}
Leung CK (2001) Delamination failure in concrete beams retrofitted with a
  bonded plate.
\newblock {\em Journal of Materials in Civil Engineering} 13(2):106--113.

\bibitem{buyukozturk2004progress}
Buyukozturk O, Gunes O, Karaca E (2004) Progress on understanding debonding
  problems in reinforced concrete and steel members strengthened using frp
  composites.
\newblock {\em Construction and Building Materials} 18(1):9--19.

\bibitem{chen2010approaches}
Chen J, Bull S (2010) Approaches to investigate delamination and interfacial
  toughness in coated systems: an overview.
\newblock {\em Journal of Physics D: Applied Physics} 44(3):034001.

\bibitem{buyukozturk2011structural}
B{\"u}y{\"u}k{\"o}zt{\"u}rk O, Buehler MJ, Lau D, Tuakta C (2011) Structural
  solution using molecular dynamics: Fundamentals and a case study of
  epoxy-silica interface.
\newblock {\em International Journal of Solids and Structures}
  48(14-15):2131--2140.

\bibitem{li2022nonlinear}
Li J, et~al. (2022) Nonlinear inclusion theory with application to the growth
  and morphogenesis of a confined body.
\newblock {\em Journal of the Mechanics and Physics of Solids} 159:104709.

\bibitem{zhang2021morphogenesis}
Zhang Q, et~al. (2021) Morphogenesis and cell ordering in confined bacterial
  biofilms.
\newblock {\em Proceedings of the National Academy of Sciences} 118(31).

\bibitem{knowles1965finite}
Knowles JK, Jakub MT (1965) Finite dynamic deformations of an incompressible
  elastic medium containing a spherical cavity.
\newblock {\em Archive for Rational Mechanics and Analysis} 18(5):367--378.

\bibitem{cohen2015dynamic}
Cohen T, Molinari A (2015) Dynamic cavitation and relaxation in incompressible
  nonlinear viscoelastic solids.
\newblock {\em International Journal of Solids and Structures} 69:544--552.

\bibitem{henzel2022reciprocal}
Henzel T, Senthilnathan C, Cohen T (2022) A reciprocal theorem for finite
  deformations in incompressible bodies.
\newblock {\em arXiv preprint arXiv:2201.08338}.

\bibitem{durban1997spherical}
Durban D, Fleck N (1997) Spherical cavity expansion in a drucker-prager solid.
\newblock {\em Transactions of the ASME: Journal of Applied Mechanics}
  64:743--750.

\bibitem{lakrout1999direct}
Lakrout H, Sergot P, Creton C (1999) Direct observation of cavitation and
  fibrillation in a probe tack experiment on model acrylic
  pressure-sensitive-adhesives.
\newblock {\em The Journal of Adhesion} 69(3-4):307--359.

\bibitem{zosel1989adhesive}
Zosel A (1989) Adhesive failure and deformation behaviour of polymers.
\newblock {\em The journal of adhesion} 30(1-4):135--149.

\bibitem{kirketerp2008distribution}
Kirketerp-M{\o}ller K, et~al. (2008) Distribution, organization, and ecology of
  bacteria in chronic wounds.
\newblock {\em Journal of clinical microbiology} 46(8):2717--2722.

\bibitem{gjodsbol2006multiple}
Gj{\o}dsb{\o}l K, et~al. (2006) Multiple bacterial species reside in chronic
  wounds: a longitudinal study.
\newblock {\em International wound journal} 3(3):225--231.

\bibitem{buret1991vivo}
Buret A, Ward K, Olson M, Costerton J (1991) An in vivo model to study the
  pathobiology of infectious biofilms on biomaterial surfaces.
\newblock {\em Journal of biomedical materials research} 25(7):865--874.

\bibitem{costerton1999bacterial}
Costerton JW, Stewart PS, Greenberg EP (1999) Bacterial biofilms: a common
  cause of persistent infections.
\newblock {\em science} 284(5418):1318--1322.

\bibitem{rybtke2015pseudomonas}
Rybtke M, Hultqvist LD, Givskov M, Tolker-Nielsen T (2015) Pseudomonas
  aeruginosa biofilm infections: community structure, antimicrobial tolerance
  and immune response.
\newblock {\em Journal of molecular biology} 427(23):3628--3645.

\bibitem{garrett2008bacterial}
Garrett TR, Bhakoo M, Zhang Z (2008) Bacterial adhesion and biofilms on
  surfaces.
\newblock {\em Progress in Natural Science} 18(9):1049--1056.

\bibitem{seminara2012osmotic}
Seminara A, et~al. (2012) Osmotic spreading of bacillus subtilis biofilms
  driven by an extracellular matrix.
\newblock {\em Proceedings of the National Academy of Sciences}
  109(4):1116--1121.

\bibitem{fei2020nonuniform}
Fei C, et~al. (2020) Nonuniform growth and surface friction determine bacterial
  biofilm morphology on soft substrates.
\newblock {\em Proceedings of the National Academy of Sciences}
  117(14):7622--7632.

\bibitem{song2015effects}
Song F, Koo H, Ren D (2015) Effects of material properties on bacterial
  adhesion and biofilm formation.
\newblock {\em Journal of dental research} 94(8):1027--1034.

\bibitem{Zhange2107107118}
Zhang Q, et~al. (2021) Morphogenesis and cell ordering in confined bacterial
  biofilms.
\newblock {\em Proceedings of the National Academy of Sciences} 118(31).

\bibitem{bjarnsholt2013vivo}
Bjarnsholt T, et~al. (2013) The in vivo biofilm.
\newblock {\em Trends in microbiology} 21(9):466--474.

\bibitem{yan2018bacterial}
Yan J, et~al. (2018) Bacterial biofilm material properties enable removal and
  transfer by capillary peeling.
\newblock {\em Advanced Materials} 30(46):1804153.

\end{thebibliography}
\end{document}


\maketitle
\thispagestyle{firststyle}
\ifthenelse{\boolean{shortarticle}}{\ifthenelse{\boolean{singlecolumn}}{\abscontentformatted}{\abscontent}}{}

\vspace{-20mm}

\newcounter{defcounter}
\setcounter{defcounter}{0}
\newenvironment{myequation}{%
        \addtocounter{equation}{-1}
        \refstepcounter{defcounter}
        \renewcommand\theequation{S\thedefcounter}
        \begin{equation}}
{\end{equation}}

\setcounter{figure}{0}
\renewcommand{\thefigure}{S\arabic{figure}}

\setcounter{subsection}{0}
\renewcommand{\thesubsection}{S.\arabic{subsection}}

\setcounter{section}{0}
\renewcommand{\thesection}{S.\arabic{section}}

\setcounter{section}{0}
\renewcommand{\thetable}{S.\arabic{table}}

\section{Pressurized Interfacial Failure (PIF) Experimental Method}
The PIF method was proposed by Wahdat et al. in {\cite{wahdat2022pressurized}}. Here we briefly describe the experimental system and its translation to compare with the theoretical predictions on this work.   

\begin{figure}[h]
\centering
\includegraphics[width=0.65\linewidth]{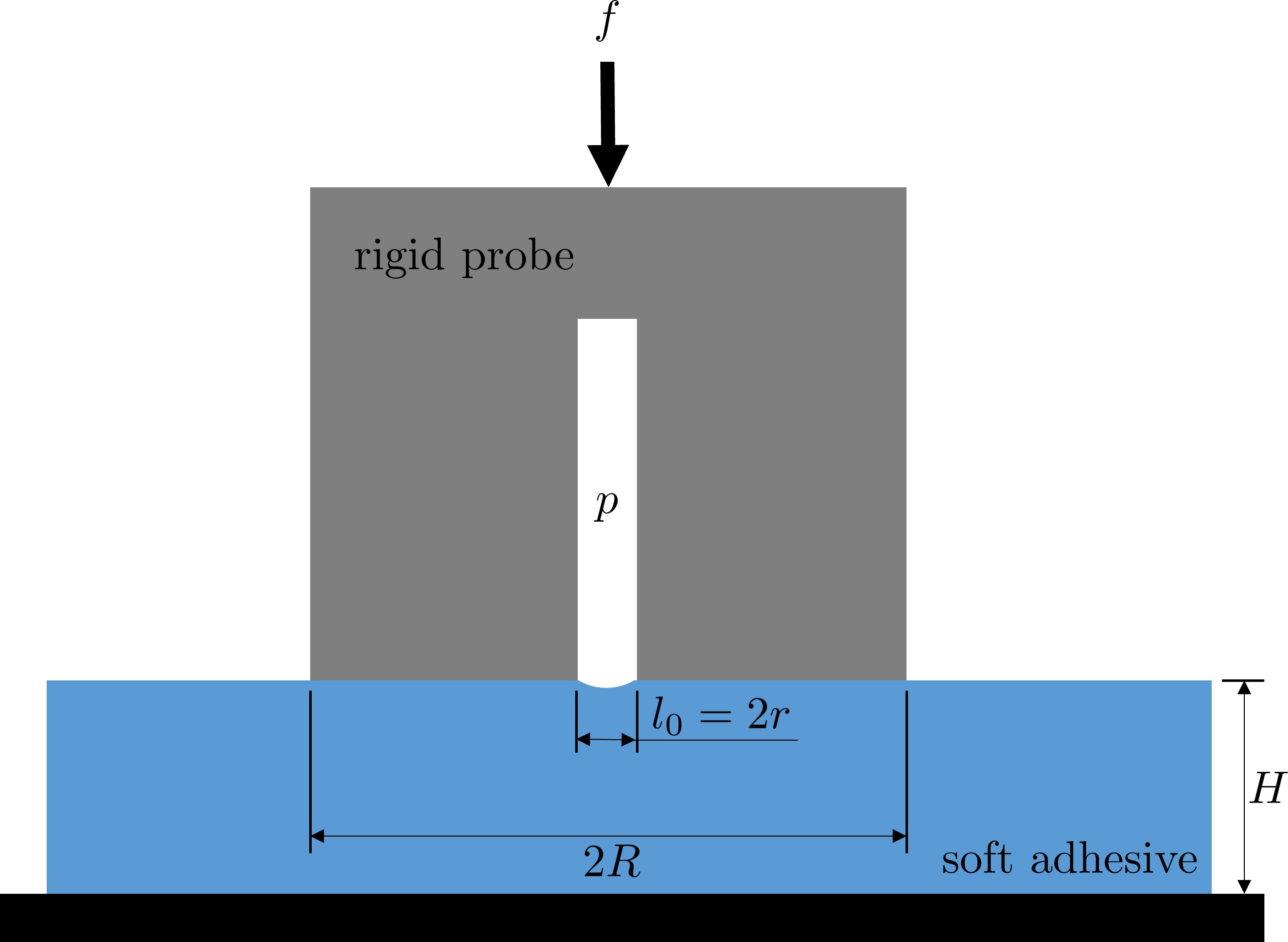}
\caption{Schematic illustration of the PIF set-up.}
\label{probe}
\end{figure}

A rigid annular probe with outer diameter $R$ and inner diameter $r$ initially pushes against an adhesive layer of thickness $H$, leaving a circular unadhered region at the interface that is enclosed by a much larger adhered region, as illustrated in Fig. \ref{probe}. Ideally, for the purpose of the present work, the dimensions are such that $r/R,r/H\ll 1$, and the applied compressive force, $f$, translates to an approximately uniform hydrostatic stress field $\sigma_h=f/\pi R^2$ in the region beneath the probe, if the initial internal pressure of the probe, $p$, is set equal to the applied hydrostatic stress (i.e. $p=\sigma$). In our set-up we have $R=1000\mu$m and $r=100\mu$m. Our adhesive layers are fabricated with $h=1000\mu$m or $300\mu$m. The initial compression is chosen to ensure that full adhesive contact is achieved. The displacement of the probe is then fixed and the internal pressure $(p)$ is gradually increased by   compressing the gas in the volume enclosed by the inner walls of the probe and the sample. The critical pressure, $\hat p_c$, is identified once delamination initiates. In our experimental systems this  can be directly inferred from the change in force $f$ that is measured simultaneously. As explained in the main text and in \cite{henzel2022reciprocal}, the initially applied hydrostatic stress has an additive effect on the applied  critical pressure, and thus to compare with predictions in Fig. 2 of the main text we write 
\begin{equation}
p_c=\hat p_c-\sigma \label{eq}
\end{equation} 
and the initial defect length is $l_0=2r$.

\pagebreak 

\setlength{\topmargin}{-0.5in}

\section{Bulk and Interfacial Properties of Adhesives Used in PIF Experiments}
In the PIF experiments we use two material sytems:

\begin{itemize}
\item[(1)] Poly(n-butyl)acrylate (PBA) networks cross-linked by ethylene glycol dimethacrylate (EGDMA). We denote by $\phi$ the weight fraction of the total monomer mass of the EGDMA cross-linker. All PBA samples were prepared from UV-curable formulations. The preparation protocol is described in{\cite{wahdat2022pressurized}}.  
\item[(2)] Commercially available VHB Tape (3M, VHB4910, thickness = 1 mm). 
\end{itemize}
The mechanical properties of these materials have been reported in\cite{wahdat2022pressurized} and included here for completeness (Table \ref{table1}). The shear modulus, $\mu$, was inferred from force displacement curves in the initial compression of the sample  with the PIF probe. The interfacial toughnesses were inferred from the propagation of the delamination  using the linear model described in\cite{wahdat2022pressurized} ($\Gamma_{\rm PIF}$), and from the conventional sphere-probe tack test ($\Gamma_{\rm SPT}$). All materials are assumed to be incompressible.

\renewcommand{\arraystretch}{1.4}

\begin{table}[h]
\centering
\begin{tabular}{ccccc} 

 \toprule

Material  & $\phi$ & $\mu$(kPa)  & $\Gamma_{\rm PIF}$(J/m$^{2}$) & $\Gamma_{\rm SPT}$(J/m$^{2}$)   \\ \midrule

  \renewcommand{\arraystretch}{1} 
   
PBA &0.25& 14.33(1.33) & 1.508 & 0.563  \\ 
PBA&0.25                     & 14.33(1.33) & 1.687 & 0.563 \\ 
PBA&0.25                     & 11.67(1.33) & 2.103 & 0.601 \\
PBA&0.25                    & 11.67(1.33) & 1.953 & 0.601 \\
PBA&0.5                    & 36(2.33)    & 0.728 & 0.776 \\
PBA&0.5                       & 36(2.33)    & 1.001 & 0.776 \\
PBA&0.5                      & 31.33(0.33) & 1.235 & 0.656 \\
PBA&0.5                      & 31.33(0.33) & 1.824 & 0.656 \\
PBA&1                        & 100(6.67)   & 1.743 & 0.052 \\
PBA&1                         & 300(6.67)   & 0.712 & 0.052\\
PBA&1                      & 89.33(3)    & 0.503 & 0.050 \\
PBA&1                        & 89.33(3)    & 0.342 & 0.050 \\
VHB& -                        & 27(1)       & 5.924 & 1.670 \\
VHB& -                          & 27(1)       & 5.478 & 1.670 \\ \bottomrule

\end{tabular}
\caption{Mechanical properties of adhesive layers. Values in parentheses represent the standard deviation. Additional details on the experimental methods are found in{\cite{wahdat2022pressurized}} and in the references therein.}
\label{table1}
\end{table}

\section{Critical Pressure Measurements Obtained by PIF Method}
In the present paper, to compare the PIF results with the theoretical predictions for expansion of interfacial cavities, we focus our attention to the critical pressure \eqref{eq}. The experimental data used to create Fig. 2, in the main text, is provided in Table \ref{table2}. For the comparison with the theoretical results in Fig. 2,  $\varphi$ is calculated for each sample using the interfacial toughness and shear modulus values from Table \ref{table1}, and $l_{0}$=0.2 mm.

\begin{table}[!h]
\centering
\begin{tabular}{cccccc} 

\toprule

Material  & $\phi$ & $H(\mu$m) & $\sigma$(kPa) & ${\hat p}_c$(kPa) & $p_c$(kPa)  \\ \midrule

  \renewcommand{\arraystretch}{1} 
   
PBA&0.25&300                & 4.983  & 26.090  & 21.107 \\ 
PBA&0.25&300                     & 21.068 & 41.327  & 20.259 \\ 
PBA&0.25&1000                   & 11.086 & 32.788  & 21.702 \\
PBA&0.25&1000                   & 1.694  & 25.767  & 24.073 \\ 
PBA&0.5&300                      & 13.867 & 38.705  & 24.838 \\ 
PBA&0.5&300                 & 21.392 & 46.387  & 24.995 \\ 
PBA&0.5&1000                  & 14.640 & 41.251  & 26.611 \\ 
PBA&0.5&1000                   & 22.973 & 56.556  & 33.584 \\ 
PBA&1&300                        & 95.040 & 137.620 & 42.580 \\ 
PBA&1&300                     & 30.443 & 72.941  & 42.498 \\ 
PBA&1&1000                       & 16.215 & 45.522  & 29.307 \\ 
PBA&1&1000                  & 48.002 & 66.132  & 18.130 \\ 
VHB&-&1000                      & 9.152  & 91.603  & 82.452 \\ 
VHB&-&1000                     & 54.289 & 114.927 & 60.638  \\ \bottomrule

\end{tabular}
\caption{Measured applied hydrostatic stress, and critical pressures, for samples of different material compositions and with different initial thickness.}
\label{table2}
\end{table}

\section{Experimental Details of the Biofilm Growth Experiments  }

\textbf{Sample preparation:} The bacterial strain used for biofilm experiments was a derivative of the \textit{Vibrio cholerae} strain C6706 with a point mutation in the \textit{vpvC} gene, which resulted in upregulated biofilm production. We further deleted the genes \textit{rbmA}, \textit{bap1}, and \textit{rbmC}, to exclude the effects of cell-substrate and cell-cell adhesion. To perform the experiments, the bacteria were first grown overnight in Lysogeny broth at 37°C under shaken conditions. The overnight culture was then diluted 30x into M9 media supplemented with 0.5\% glucose, 2 mM MgSO$_4$ and 100 $\mu$M CaCl$_2$ and grown at 30°C for 1-2 hours under shaken conditions. This culture was then diluted to an OD$_{600}$ of $1-3 \times 10^{-3}$, and a  1 $\mu$L droplet of the diluted solution was placed in the center of a glass-bottomed 96 well plate (MatTek). The droplet was then covered with 20 $\mu$L of molten agarose gel of concentration 0.2-1\%, encasing the cells. Upon cooling, 200 $\mu$L of the above-mentioned supplemented M9 media was added. The plates were incubated at 30°C and either imaged periodically or after 12-16 hrs.

\vspace{2mm}
\noindent\textbf{Characterization of agarose stiffness:} The stiffness of the agarose gels was measured using a shear rheometer (Anton Paar Physica). 

\vspace{2mm}
\noindent\textbf{Surface treatment:} To reduce the interfacial toughness between the agarose gel and the glass substrate, we treated the substrate using vapor deposition of silane by enclosing the 96 well plate in a box with a 1 mL reservoir of 3‐aminopropyltriethoxysilane (Sigma) for 16 hrs.

\vspace{2mm}
\noindent\textbf{Imaging:} Images were taken using a confocal spinning disk unit (Yokogawa CSU-W1; Nikon Eclipse Ti2; Photometrics Prime BSI). For time-lapse imaging, a 100x silicon oil immersion objective was used to image biofilms at z-intervals of 0.13 or 0.195 $\mu$m. For endpoint imaging of mature biofilm contact angles, a 60x water immersion objective was used to image the bottom 5 $\mu$m of the biofilms at z-intervals of 0.4 $\mu$m. In the latter case, ~100 fields of view were imaged resulting in an average of 89 (range 27-140) unique biofilm measurements for each concentration. 

\vspace{2mm}
\noindent\textbf{Image analysis:} Biofilm images were de-noised and deconvolved using Huygens SVI software and binarized using Otsu thresholding in MATLAB (2018a). For each slice in the confocal image, a convex hull, which encompassed all binarized pixels in each biofilm, was found. The area of the convex hull was used to estimate the height-dependent cross-sectional area of the biofilm $A(z)$ from which the effective radius was calculated as $r(z)=(A/\pi)^{1/2}$. The contact angle $\theta$ was then found by fitting a linear slope in the bottom 5 $\mu$m of the biofilm, $$\theta=\frac{180}{\pi}\left(\tan^{-1}\left(\frac{{\rm d}r}{{{\rm d}z}}\right)+\frac{\pi}{2}\right)~\text{[deg]}.$$ 

\begin{figure}[h]
\centering
   \includegraphics[width=1\textwidth]{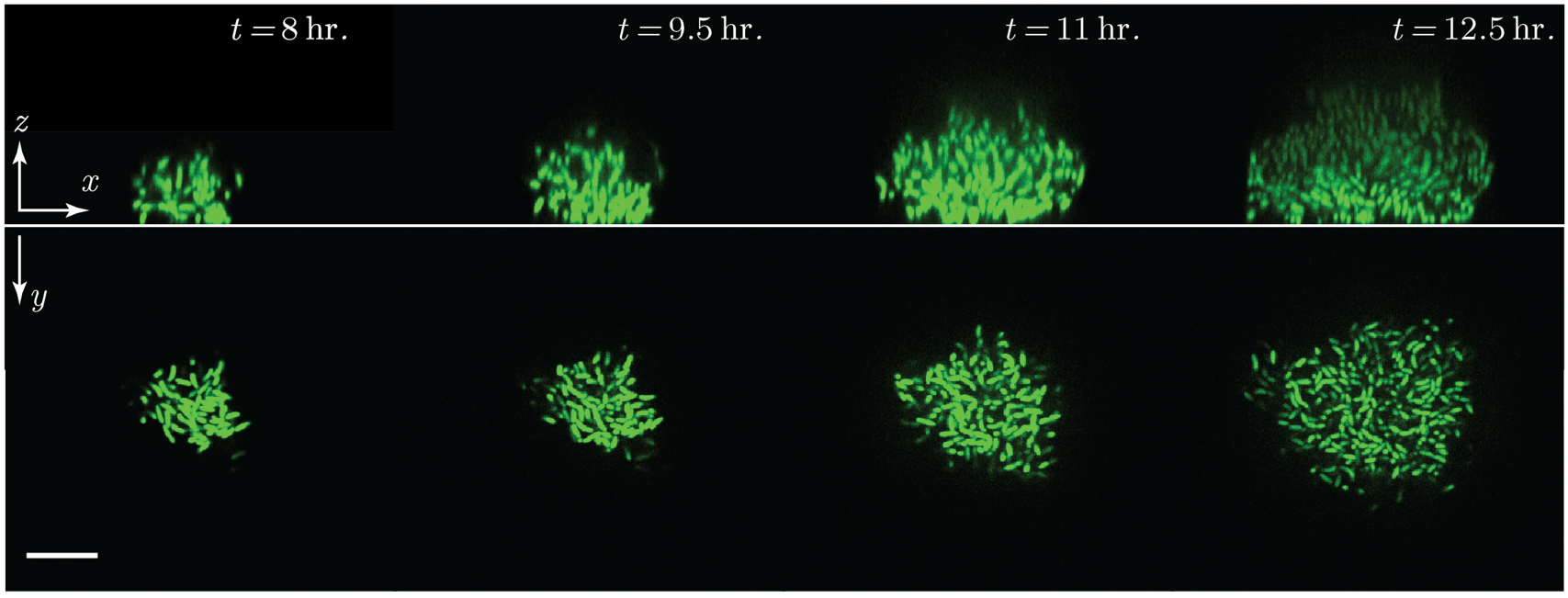}
\caption{Confocal images showing the growth of a single biofilm at different times throughout the growth process, at single cell resolution.   The biofilm is grown under the confinement of an agrose gel of stiffness $\mu$=0.16 (kPa). Scale bar is $10\mu$m.}\label{soft}
\end{figure}

\begin{figure}[h]
\centering
   \includegraphics[width=1\textwidth]{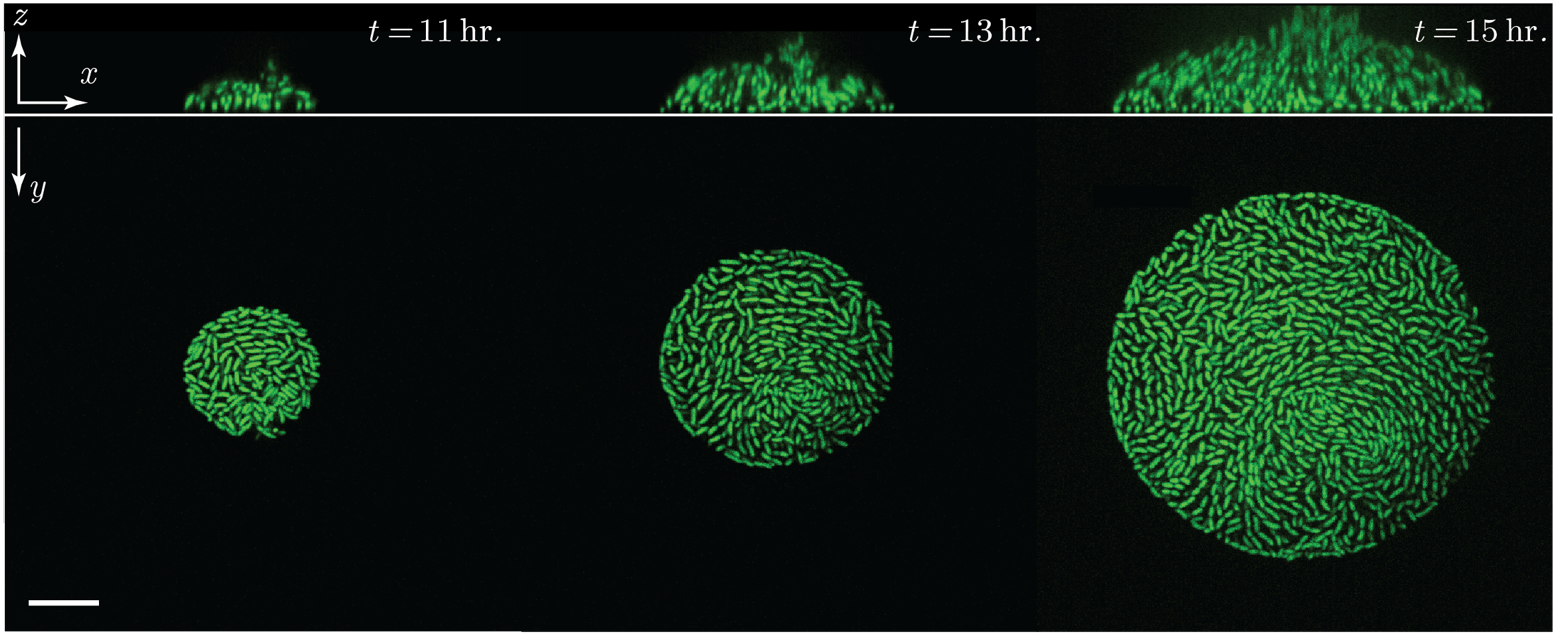}
\caption{Confocal images showing the growth of a single biofilm at different times throughout the growth process, at single cell resolution. The biofilm is grown under the confinement of an agrose gel of stiffness $\mu$=4.7 (kPa). Scale bar is $10\mu$m.}\label{stiff}
\end{figure}

\vspace{2mm}
\noindent\textbf{Biofilm growth:} Images of  biofilms at different times throughout the growth process, as referred to in Fig. 3. of the main text, can be found in  Figs. \ref{soft} and \ref{stiff}.


